\documentclass[%
 reprint,%
 amssymb, amsmath,%
 aip,cha,%
]{revtex4-1}

\usepackage{bm}%
\usepackage[colorlinks=true,linkcolor=blue]{hyperref}%
\usepackage{graphicx}
\usepackage{epstopdf}
\usepackage[T1]{fontenc}
\usepackage[latin9]{inputenc}
\usepackage{amsbsy}
\usepackage{gensymb}

\expandafter\ifx\csname package@font\endcsname\relax\else
 \expandafter\expandafter
 \expandafter\usepackage
 \expandafter\expandafter
 \expandafter{\csname package@font\endcsname}%
\fi
\begin{document}
\title{\textrm{Room Temperature Magnetoresistance and Exchange Bias in "314 - type" Oxygen-Vacancy Ordered SrCo$_{0.85}$Fe$_{0.15}$O$_{2.62}$}}%

\author{Prachi Mohanty}
\affiliation{Indian Institute of Science Education and Research Bhopal, Bhopal, 462066, India}
\author{Sourav Marik}
\email[]{soumarik@gmail.com}
\affiliation{Indian Institute of Science Education and Research Bhopal, Bhopal, 462066, India}
\author{Deepak Singh}
\affiliation{Indian Institute of Science Education and Research Bhopal, Bhopal, 462066, India}
\author{Ravi P. Singh}
\email[]{rpsingh@iiserb.ac.in}
\affiliation{Indian Institute of Science Education and Research Bhopal, Bhopal, 462066, India}

\date{May 2017}%

\begin{abstract}
Herein, we report the magneto-transport and exchange bias effect in a "314 - type" oxygen - vacancy ordered material with composition SrCo$_{0.85}$Fe$_{0.15}$O$_{2.62}$. This material exhibits a ferrimagnetic transition above room temperature, at 315 K. The negative magnetoresistance starts to appear from room temperature (-1.3 $\%$ at 295 K in 70 kOe) and reaches a sizable value of 58 $\%$ at 4 K in 70 kOe. Large exchange bias effect is observed below 315 K when the sample is cooled in the presence of a magnetic field. The coexistence of nearly compensated and ferrimagnetic regions in the layered structure originate magnetoresistance and exchange bias in this sample. The appearance of a sizable magnetoresistance and giant exchange bias effect, especially near room temperature indicates that "314-type" cobaltates are a promising class of material systems for the exploration of materials with potential applications as magnetic sensors or in the area of spintronics. 
\end{abstract}

\maketitle

	Transition metal oxides (TMOs) with strong electron correlations are a fascinating class of material systems exhibiting exotic electronic and magnetic complexity which can be turned into promising technological applications. Example includes high-T$_{c}$ superconductivity (SC) in cuprates, \cite{Bednorz} colossal magnetoresistance (CMR) in manganites \cite{von} and multiferroicity in bismuth compounds.\cite{Zhao} For the practical utilization of these properties, a transition temperature near or above room temperature (RT) is usually required. In this context, ferrimagnetic TMOs with transition higher than RT have attracted considerable interest in recent times. For instance, Sr$_{2}$FeMoO$_{6}$ and Sr$_{2}$FeReO$_{6}$ materials having ferrimagnetic transition higher than 400 K show room temperature magnetoresistance. \cite{Kobayashi,Tomioka,Kimura} Transition-metal-only ferrimagnetic double perovskites with chemical composition Mn$_{2}$BReO$_{6}$ (B = Fe and Mn) show a new way to control magnetoresistance in spintronic materials. \cite{Attfield,Retuerto,Hodges} The non-stoichiometric Ba$_{2}$Fe$_{1.12}$Os$_{0.88}$O$_{6}$ compound (ferrimagnetic transition temperature T$_{c}$ = 370 K) shows the appearance of exchange bias effect near room temperature.\cite{Feng}

In this letter, we report the room temperature exchange bias effect  and magnetoresistance for "314-type" cobaltate with composition SrCo$_{0.85}$Fe$_{0.15}$O$_{2.62}$. The oxygen deficient "314-type" compounds crystallize in a tetragonal symmetry, space group $I4/mmm$ with a 2a$_{p}$ $\times$ 2a$_{p}$ $\times$ 4a$_{p}$ (a$_{p}$ = lattice parameter of the cubic perovskite) type unit cell and show the appearance of above RT ferrimagnetism. \cite{Istomin,Li,Sheptyakov,Kishida,Nakao,Marik}

A pure phase polycrystalline sample of SrCo$_{0.85}$Fe$_{0.15}$O$_{2.62}$ was prepared by standard solid-state reaction method as reported previously. \cite{Marik} The sample was characterized by X-ray powder diffraction (XRD) at room temperature (RT), performed in a Panalytical X'pert Pro diffractometer (Cu K$_{\alpha}$ radiation, $\lambda$ = 1.5406 $\text{\AA}$). Resistances of the polycrystalline sintered bar was measured on a Quantum Design physical property measurement system between 4-350 K and in magnetic fields up to 70 kOe. Temperature and field-dependent magnetization measurements were carried out by using Quantum Design MPMS-3 magnetometer. \begin{figure}[h]
\includegraphics[width=1.0\columnwidth]{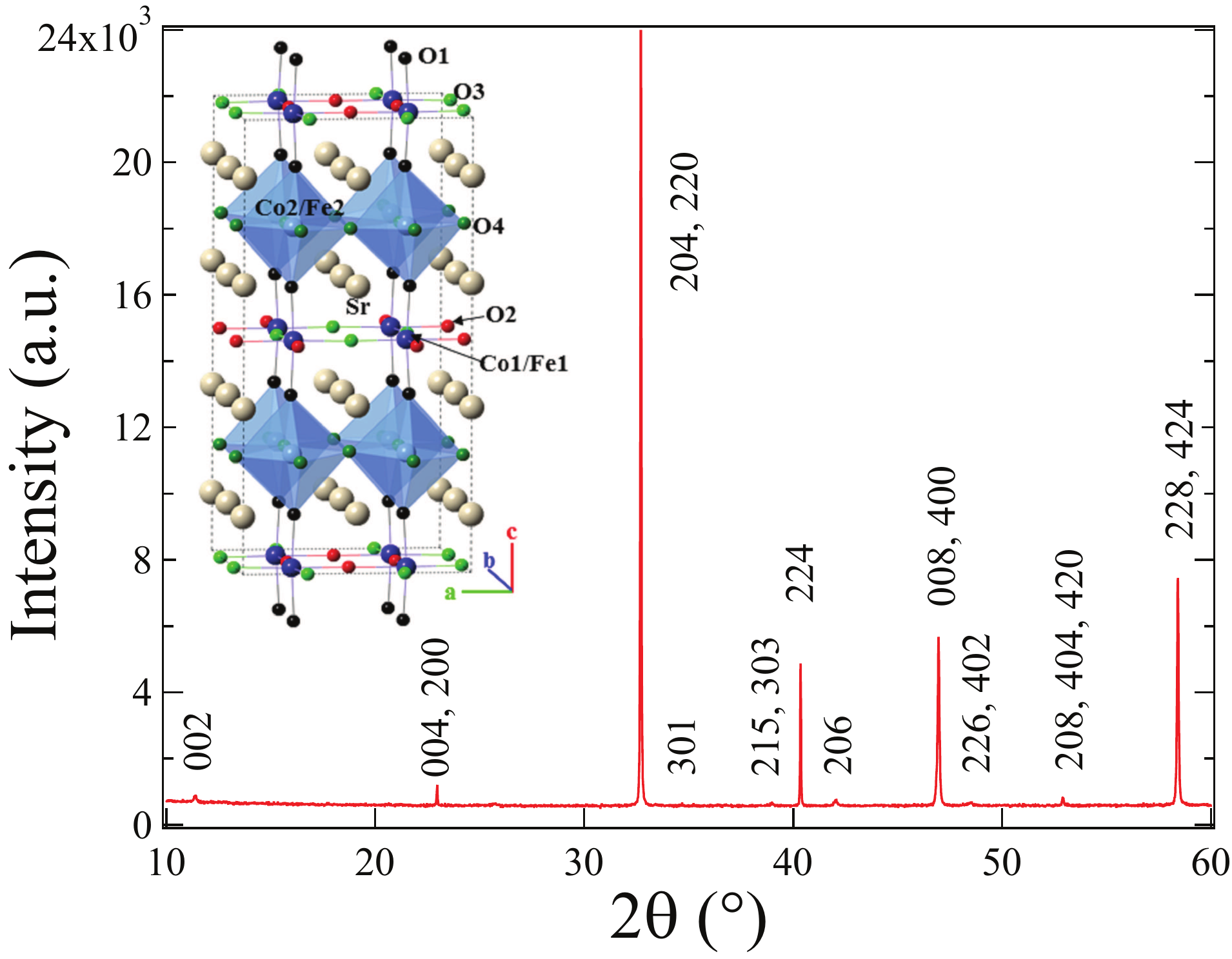}
\caption{\label{Fig1:xrd} Room temperature powder X-ray diffraction pattern of SrCo$_{0.85}$Fe$_{0.15}$O$_{2.62}$ indexed with the "314-type" tetragonal 2a$_{p}$ $\times$ 2a$_{p}$ $\times$ 4a$_{p}$ (a$_{p}$ = lattice parameter of the cubic perovskite) unit cell. The inset shows the crystal structure of the same material. }
\end{figure}

Figure 1 shows the RT-XRD pattern for the SrCo$_{0.85}$Fe$_{0.15}$O$_{2.62}$ sample, indexed with the "314-type" tetragonal cell. The crystal structure is shown in the inset of figure 1. As suggested previously, \cite{Marik} the oxygen vacancy ordering in the oxygen-deficient Co1/Fe1-O layers and the tilting of octahedra (Co1/Fe1-O1-Co2/Fe2 = 172.9$\degree$, see ref. 16) due to the Jahn-Teller 
distortion of the IS Co$^{3+}$ (t$_{2g}^{5}$e$_{g}^{1}$, S = 1) in the oxygen-replete region trigger the quadrupling of the c axis with respect to the oxygen vacancy disordered cubic cell.
	
Figure 2 shows the temperature variation of the FC susceptibility (measured with 1 kOe magnetic field), which confirms the ferrimagnetic transition at 315 K. Nevertheless, the increase in the $\chi_{FC}$ at lower temperature can be attributed to the frustration in the layered structure due to the existence of antiferromagnetic (AFM, Fe$^{4+}$-O(2p)-Co$^{3+}$ and Fe$^{4+}$-O(2p)-Fe$^{4+}$) and ferromagnetic (FM, Fe$^{4+}$-O(2p)-Co$^{4+}$, Co$^{4+}$-O(2p)-Co$^{4+}$ and Co$^{3+}$-O(2p)-Co$^{4+}$) interactions.
\begin{figure}
\includegraphics[width=1.0\columnwidth]{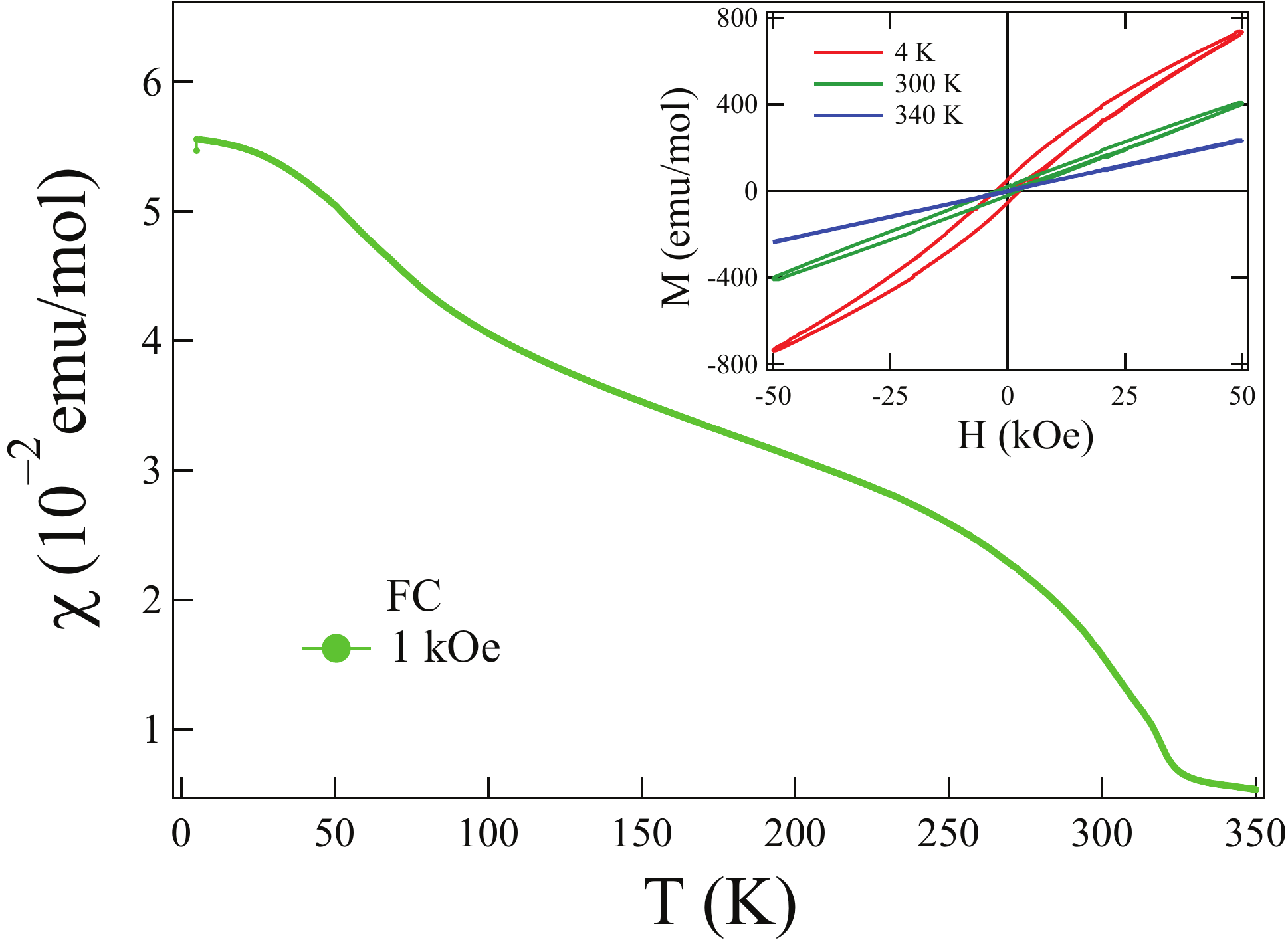}
\caption{\label{Fig2:MT} Temperature variation of field cooled (FC, Magnetic field = 1 kOe) magnetic susceptibility ($\chi$) for SrCo$_{0.85}$Fe$_{0.15}$O$_{2.62}$. The inset shows the zero-field-cooled magnetic hysteresis measurements (M-H) at several temperatures.}
\end{figure}

Temperature variation of the electrical resistivity at 0 kOe ($\rho(T)_{H=0 kOe}$) and 70 kOe ($\rho(T)_{H=70 kOe}$) magnetic field exhibit semiconducting-type behavior and a change in slope are observed at the ferrimagnetic transition temperature ($\sim$ 315 K, left inset in figure 3). The resistivity data measured at 70 kOe do not show any change up to 300 K however, below 300 K $\rho(T)_{H=70 kOe}$ starts to deviate from $\rho(T)_{H=0 kOe}$ and becomes smaller down to 4 K. The temperature variation of magnetoresistance (MR), which is defined as MR = [$\rho(T)_{H}$- $\rho(T)_{H=0 kOe}$]/$\rho(T)_{H=0 kOe}$ is shown in figure 3. Interestingly, a sizeable negative MR is observed over a wide temperature range (300-4K), and at 4 K the MR is -56$\%$. In order to confirm the magnetoresistance behavior of the investigated sample, magnetic field dependent resistances ($\rho(H)$) were measured. The sample was cooled down (with zero magnetic field) from T = 350 K to the measurement temperatures and then measurements were performed. The right inset in figure 3 shows field variation of the MR at several temperatures. The MR values monotonically increase with increasing magnetic field and the highest MR value at 70 kOe reaches 58 $\%$ at 4 K. It is noteworthy that in contrast with perovskite manganites, only a handful of cobalt based perovskite compounds show MR. For instance, the oxygen deficient perovskite SrCoO$_{2.75}$ \cite{Taguchi,Maignan} shows a negligible MR value of 0.5 $\%$ in 70 kOe, the MR was not reported both for Sr-rich La$_{1-x}$Sr$_{x}$CoO$_{3- \delta}$ (0.5 $\le$ x $\le$ 0.9) \cite{Sunstrom} and electrochemically oxidized \cite{Bezdicka} or high-pressure synthesized \cite{Takeda} SrCoO$_{3}$. In fact, only two series of Co-based perovskites, La-rich La$_{1-x}$Sr$_{x}$CoO$_{3}$ (x $<$ 0.5) \cite{Briceno,Yamaguchi,Mahendiran} and the ordered oxygen deficient $Ln$BaCo$_{2}$O$_{5.4}$ ($Ln$ = rare-earth elements)\cite{Martin} show a significant MR ranges from 12.5 $\%$ to 50 $\%$ in H = 70 kOe at low temperatures. Therefore, the observed MR in the present sample is higher than the two other magnetoresistive cobalt perovskite systems.\\
\begin{figure}
\includegraphics[width=1.0\columnwidth]{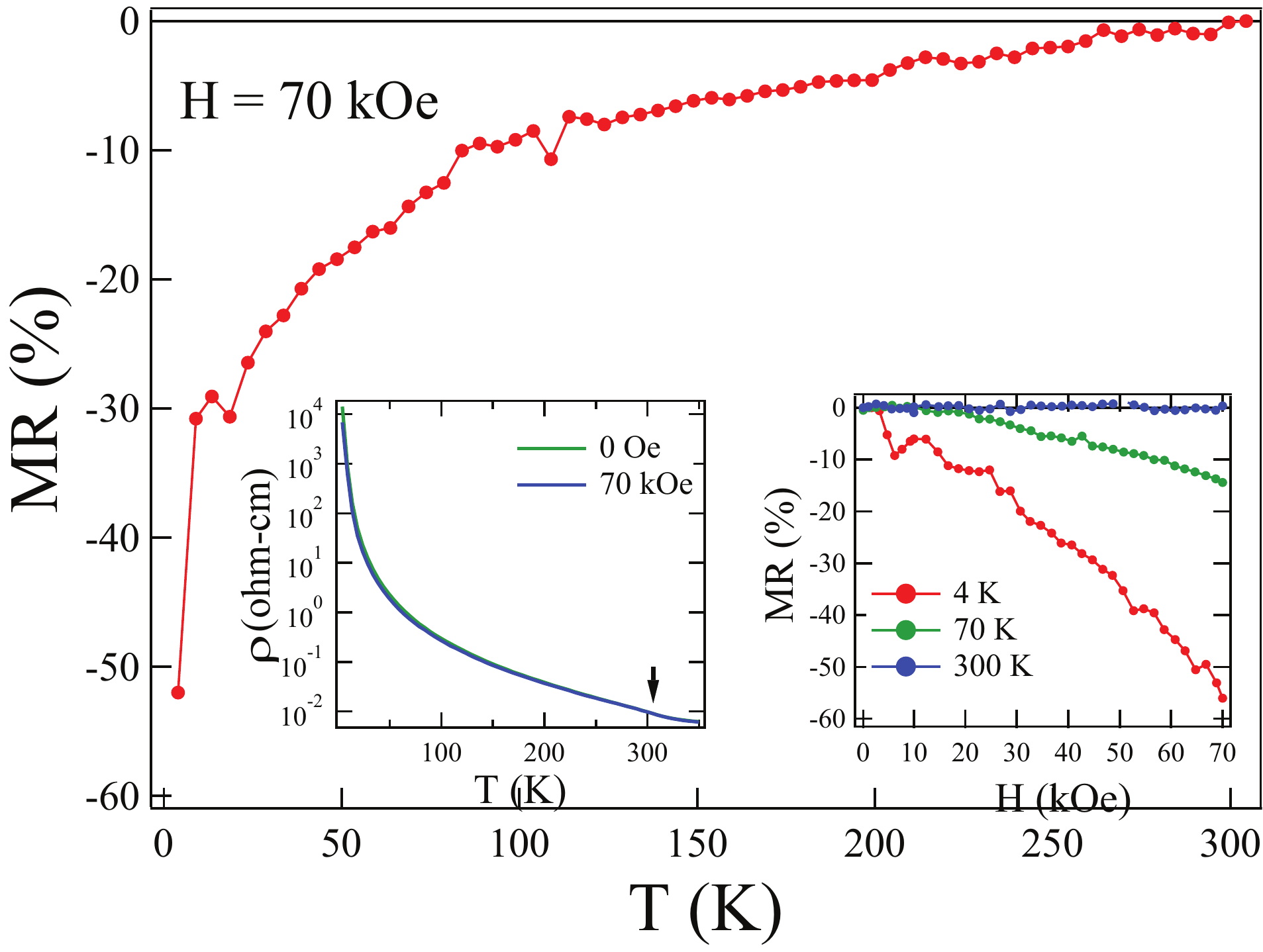}
\caption{\label{Fig3:TDR} Temperature variation of the magnetoresistance (MR = [$\rho(T)_{H}$- $\rho(T)_{H=0 kOe}$]/$\rho(T)_{H=0 kOe}$ for the SrCo$_{0.85}$Fe$_{0.15}$O$_{2.62}$ sample. The left inset shows temperature variation of the resistivity measured under 0 ($\rho(T)_{H=0 kOe}$) and 70 kOe ($\rho(T)_{H=70 kOe}$) magnetic field. A change in slope of the resistivity curve is shown by an arrow. The right inset shows a magnetic field (H) variation of the MR at several temperatures.}
\end{figure}
To understand the conduction mechanism (4 K-350 K), the resistivity data measured under zero magnetic field ($\rho(T)_{H=0 kOe}$) is analyzed by fitting (not shown) to Arrhenius-type behavior and Mott variable range hopping (VRH) type behavior expected for a three-dimensional (3D) disordered system ($\rho(T)$ = $\rho(T)_{0}$exp($T_{0}/T)^{1/(d+1)}$, for the Arrhenius model d = 0, T$_{0}$ is the activation energy and for the 3D - VRH model d = 3 and T$_{0}$ is the characteristic energy for hopping). The Arrhenius model does not follow a straight line; however, the fitting with the 3D - VRH model is more linear. Therefore, the electron conduction in SrCo$_{0.85}$Fe$_{0.15}$O$_{2.62}$ occurs by Mott 3D - VRH mechanism with the carriers localized due to the disorder. The existence of ferrimagnetism and different magnetic interactions (such as Fe$^{4+}$-O(2p)-Co$^{3+}$ and Fe$^{4+}$-O(2p)-Fe$^{4+}$ are AFM and Fe$^{4+}$-O(2p)-Co$^{4+}$, Co$^{4+}$-O(2p)-Co$^{4+}$ and Co$^{3+}$-O(2p)-Co$^{4+}$ are FM) and their competition, which can develop the spin-dependent scattering of carriers could be the origin of the MR in this sample. The magnetic disorder increases the spin-dependent scattering of the carriers at a lower temperature and thus increases the electrical resistivity in the absence of magnetic field. Importantly, the existence of Jahn-Teller active intermediate spin states of Co$^{3+}$ (t$_{2g}^{5}$e$_{g}^{1}$, S = 1) in this sample originates distortion in the octahedral local environment. Therefore, similarly to the manganites, these lattice deformations (Jahn-Teller polarons) \cite{Keller} may have an important contribution to the magnetotransport properties of this sample.

Also, in recent times the ferrimagnetic materials have attracted immense interest for the exchange bias behavior. \cite{Feng,Nayak,Sahoo} The exchange bias (EB) phenomenon is generally demonstrated as a shift \cite{Meiklejohn} in the isothermal magnetization loop with respect to horizontal magnetic field axis and vertical magnetization axis and is exploited in several technological applications such as magnetic recording read heads, \cite{Tsang} random access memories \cite{Prejbeanu} and other spintronic devices. \cite{Chen,Fuke} Figure 4(a) illustrates the magnetic hysteresis measurement (M-H) between $\pm$ 70 kOe for the SrCo$_{0.85}$Fe$_{0.15}$O$_{2.62}$ sample at 10 K, performed in field cooled (FC) method (cooling field, H$_{FC}$ = 20 kOe). For the FC M-H measurement, performed at 10 K, the sample was cooled from T = 350 K $>$T$_{C}$ $\approx$ 315 K. A clear shift of the FC M-H loop towards both the negative field and the positive magnetization direction is observed. This shift can be characterized in terms of exchange bias field (H$_{EB}$ = -(H$_{C(L)}$ + H$_{C(R)}$)/2, where H$_{C(L)}$ and H$_{C(R)}$ are the left and right intercepts of the magnetization curve with the field axis. A large value of H$_{EB}$ $\sim$ 7.35 kOe is observed at 10 K. In order to choose a suitable cooling field, we have measured the M-H loops in various cooling fields (H$_{FC}$) ranging from 0.5 kOe to 50 kOe at 10 K. As shown in the inset of figure 4(a) the exchange bias field (H$_{EB}$) shows a saturation tendency above 10 kOe.\\
\begin{figure}[h]
\includegraphics[width=1.0\columnwidth]{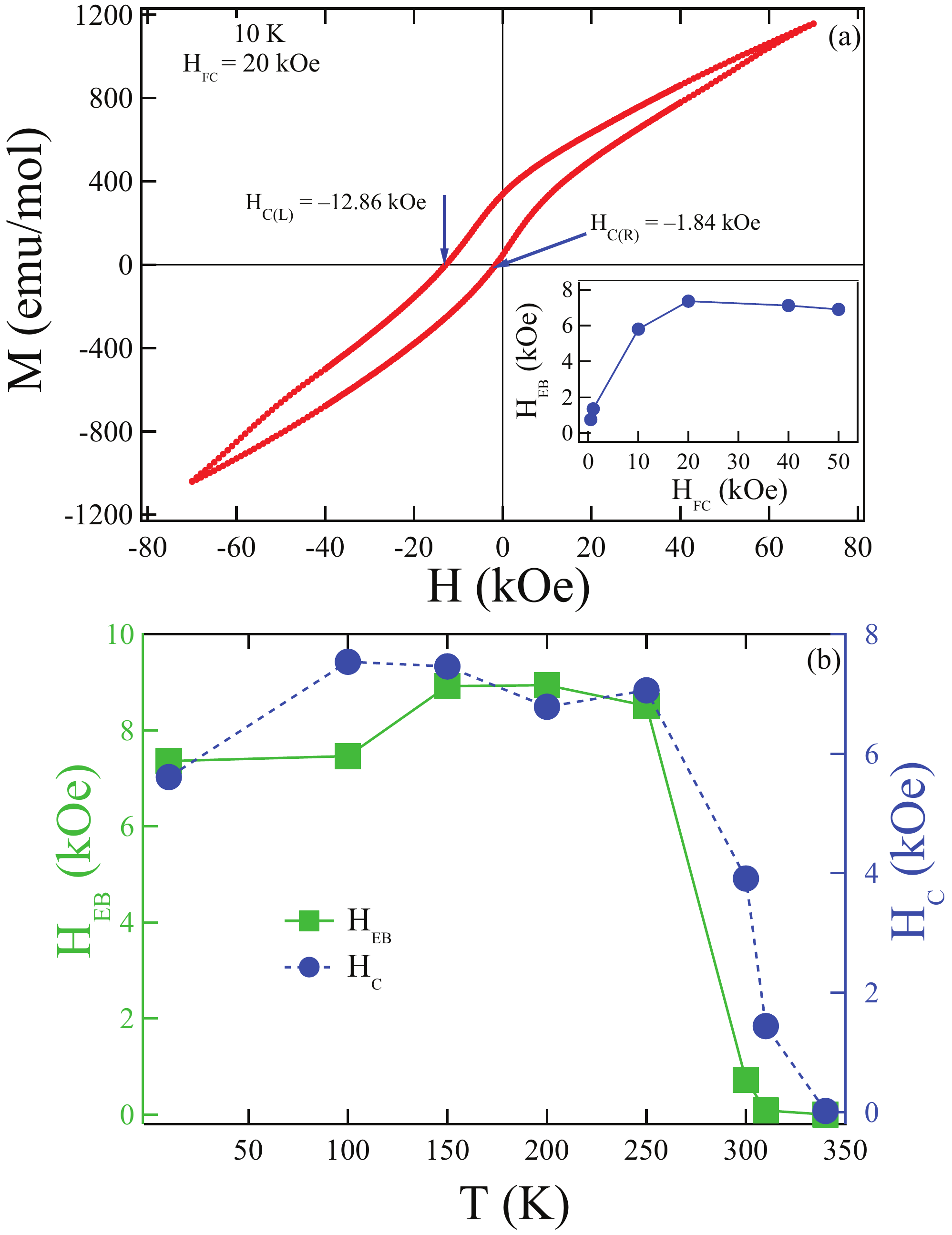}
\caption{\label{Fig4:MH} (a) Magnetic field dependent magnetization loop (M-H) of SrCo$_{0.85}$Fe$_{0.15}$O$_{2.62}$ at 10 K measured in field cooling mode (FC, cooling field H$_{FC}$ = 20 kOe). The inset shows cooling field (H$_{FC}$) dependence of the H$_{EB}$ measured at 10 K. (b) Temperature dependence (left axis) of H$_{EB}$ and (right axis) coercivity for the same.}
\end{figure}
To explore the temperature dependence of EB properties, and to verify whether the observed EB properties are associated with the magnetic ordering, we have studied the temperature dependence of exchange bias. The sample was field-cooled (FC) to the measuring temperatures in an applied field of 20 kOe. Once the measuring temperature was reached, the M-H loops were measured between $\pm$ 70 kOe. Figure 4(b) shows the temperature variation of the exchange bias field (H$_{EB}$) and coercivity (H$_{C}$). The EB effect is started $\sim$ 315 K (H$_{EB}$ = 84 Oe at 310 K) and with lowering the temperature H$_{EB}$ continuously increases up to 250 K (H$_{EB}$ = 8.5 kOe at 250 K) then shows a near saturation down to 10 K. Interestingly, the coercivity also persists up to 315 K, which is in fact, the ferrimagnetic ordering temperature for SrCo$_{0.85}$Fe$_{0.15}$O$_{2.62}$. 

Training effect is an important characteristic related to the EB property, where the uncompensated spin configuration at the interface may relax due to the repetition of the M-H loop and the H$_{EB}$ depends on the number of successive hysteresis loops measured. Figure 5 shows the variation of H$_{EB}$ with a number of field cycles (n). An enlarged view of the low field region of successive M-H loops (the left intercept of magnetic field axis) at 10 K is shown in the inset of figure 5. It clearly shows that the training effect is present in this sample and the exchange bias decreases monotonically as the cycle number (n) increases. A power law relationship is usually suggested to describe the number of field cycle (n) dependence of exchange bias. \cite{Nogu}

\begin{equation}
H_{EB} (n) - H_{E\infty} = K/\sqrt{n}  ,
\label{eqn1:l}
\end{equation}
                                                                                                        
Where H$_{EB}$ (n) is the exchange field for the nth cycle. H$_{E\infty}$ is the exchange field for n = $\infty$, and K is a system dependent constant. The fitted data (dotted line) is shown in figure 5 and the obtained value of H$_{E\infty}$ is 6.5 kOe, which will be the remnant H$_{EB}$ in the sample. Also, our experimental data matched well with the recursive formula suggested by Binek \cite{Binek} i.e., 
\begin{equation}
H_{EB} (n+1) - H_{EB} (n) = -\gamma  (H_{EB} (n) - H_{E\infty})^{3} ,
\label{eqn2:d}
\end{equation}                                                                       
where $\gamma$ = 1/2K$^{2}$.

\begin{figure}
\includegraphics[width=1.0\columnwidth]{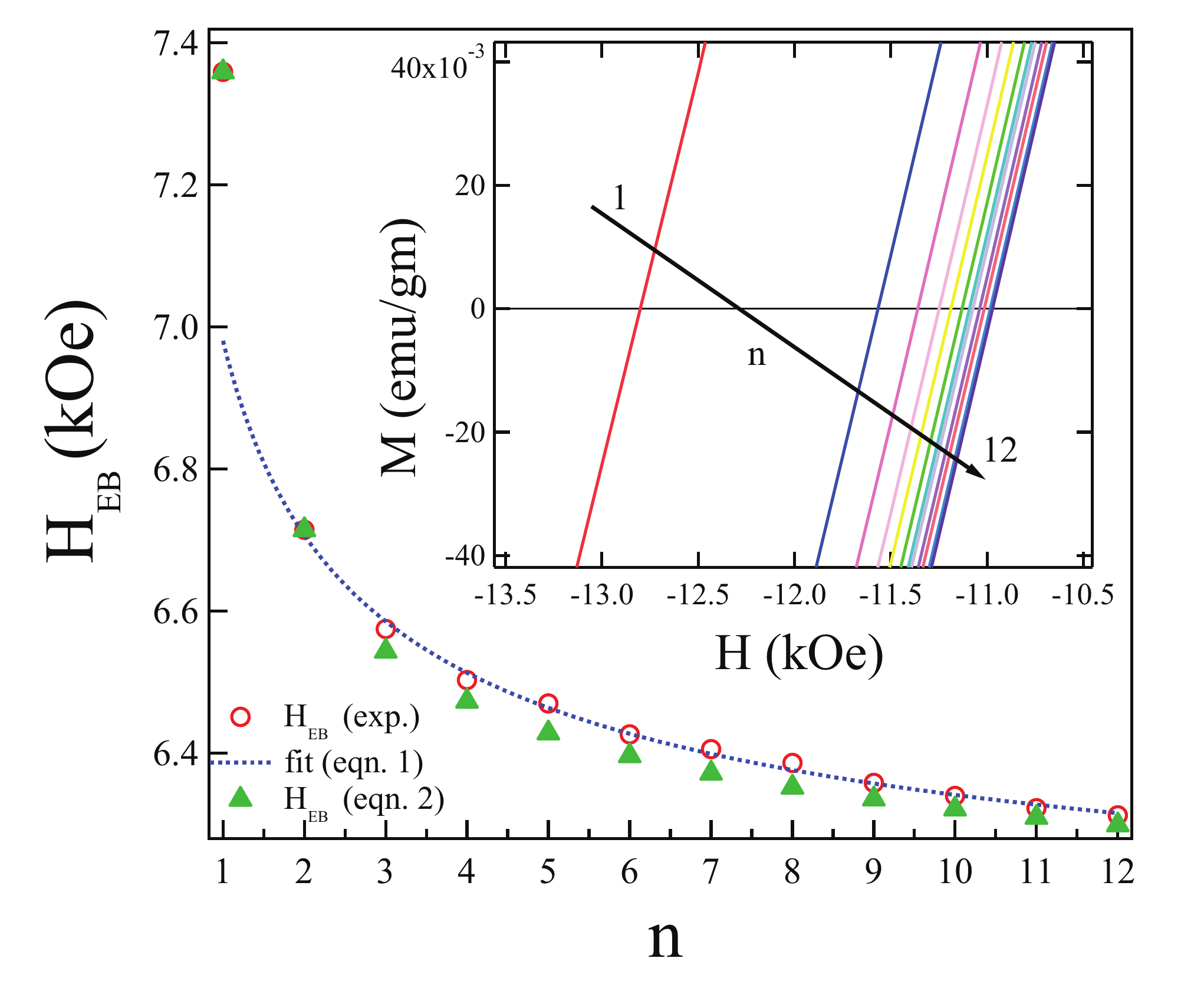}
\caption{\label{Fig5:EB} The exchange bias field (H$_{EB}$) vs number of field cycle (n) (red open circles) obtained
from the training effect magnetic hysteresis loops (M-H) at 10 K. The inset shows an enlarged view of the consecutive M-H loops (left intercept with the field axis). Arrow indicates the direction of increase in field cycle (n).}
\end{figure}

\begin{figure}
\includegraphics[width=1.0\columnwidth]{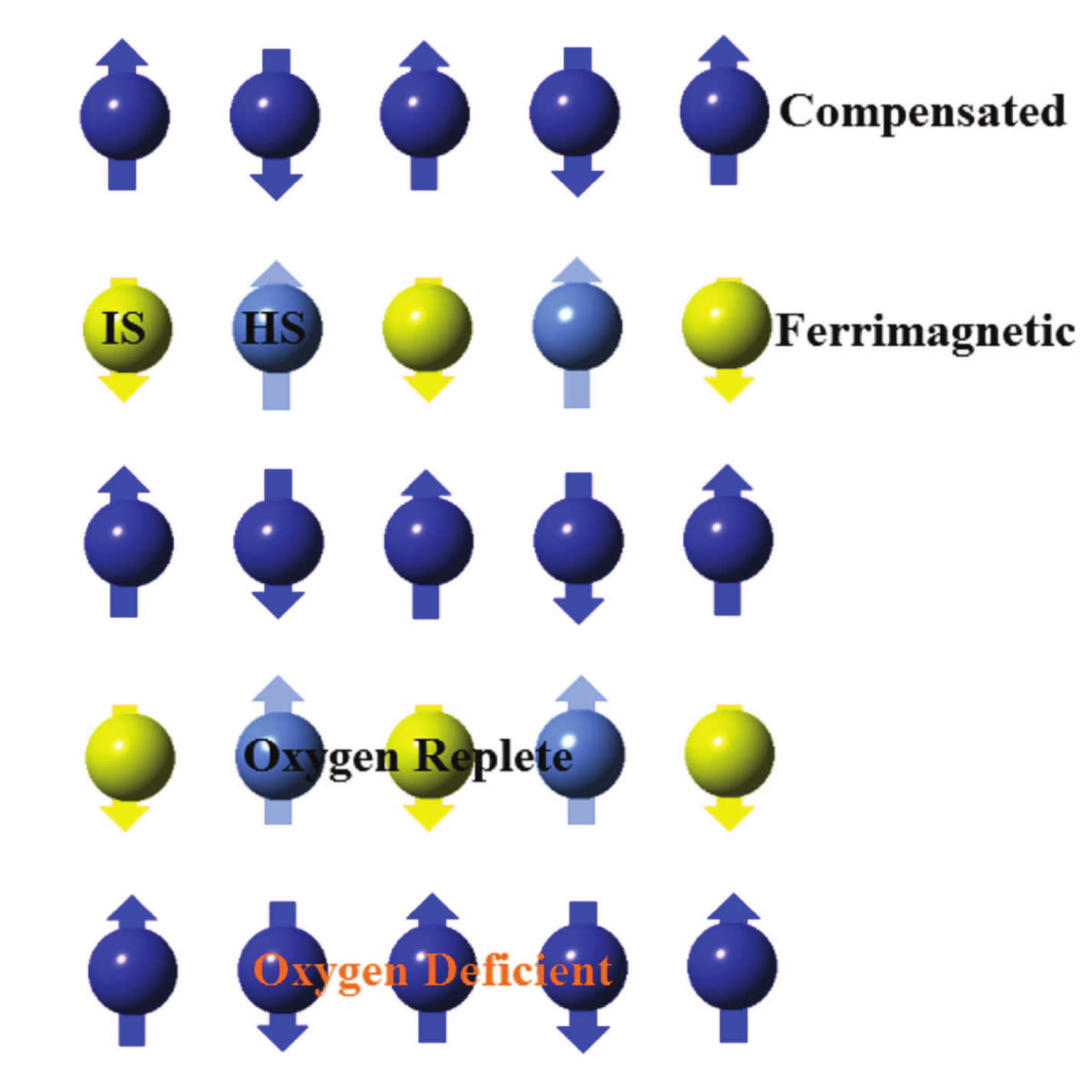}
\caption{\label{Fig6:MO} The magnetic ordering scheme of Co cations along [100] direction for SrCo$_{0.85}$Fe$_{0.15}$O$_{2.62}$ (for "314 - type" cobaltates).}
\end{figure}
Recently, giant EB effects are observed in nearly compensated ferrimagnetic systems such as, in Heusler alloys Mn$_{3-x}$Pt$_{x}$Ga \cite{Nayak} and in double perovskite Ba$_{2}$Fe$_{1.12}$Os$_{0.88}$O$_{6}$. \cite{Feng} The EB effects in those systems were attributed to the presence of ferrimagnetic clusters in the compensated host, which are formed as a consequence of antisite disorder. The magnetic ordering scheme of Co cations for the present sample (and/or any "314 - type" cobaltates, see ref. 14 and 15) is shown in figure 6. There are two different Co sites in this layered structure, one is oxygen-replete (Co2/Fe2, see figure 1) and another oxygen-deficient (Co1/Fe1). As mentioned previously\cite{Kishida, Nakao, Marik}, the ferrimagnetic structure is formed with a Co$^{3+}$ high - spin (HS, t$_{2g}^{4}$e$_{g}^{2}$, S = 2) state and intermediate-spin (IS, t$_{2g}^{5}$e$_{g}^{1}$, S = 1) state ordering (AFM aligned) in the oxygen-replete layers, while the Co cations in the oxygen deficient site are nearly compensated. Therefore, the large EB effect in this sample can be attributed to the coexistence of nearly compensated and ferrimagnetic regions. However, the existence of different magnetic interactions in the layered structure may also have an effect on the EB at a lower temperature. 
For technical applications of the EB effect, heterostructures are indeed required while the EB generally occurs at the interface of composite FM/AFM materials. Therefore, the layered structure type of the present material in combination with the potential for showing giant EB, sizable magnetoresistance and magnetic ordering above room temperature indicates that "314 - type" cobaltates are a promising class of compounds for designing new magnetic materials. 

In summary, we have shown near room temperature appearance of two technologically important phenomena namely, negative magnetoresistance and EB in a "314 - type" oxygen vacancy ordered cobaltate sample. SrCo$_{0.85}$Fe$_{0.15}$O$_{2.62}$ shows ferrimagnetic transition above RT. The negative magnetoresistance starts to appear from room temperature and reaches a sizable value of 58 $\%$ at 4 K in 7 T. The exchange bias is observed below 315 K when the sample is cooled in the presence of a magnetic field. The large EB effect in this sample is attributed to the coexistence of nearly compensated and ferrimagnetic regions. Appropriate chemical doping should result in a further increase of the magnetoresistance and exchange bias. Moreover, the appearance of a sizable MR and EB near room temperature suggest that further studies on this material and on other "314 - type" compounds could be useful for finding materials with potential applications as sensors or in the area of spintronics.\\ 

R. P. S.  acknowledges Science and Engineering Research Board (SERB), Government of India for the Ramanujan Fellowship through Grant No. SR/S2/RJN-83/2012. S. M acknowledges Science and Engineering Research Board, Government of India for the NPDF fellowship (Ref. No. PDF/2016/000348).

PM and SM have equal contributions.

\end{document}